\documentclass{elsart}

\begin{document}

\begin{frontmatter}
\title{Dynamical Origin of Decoherence in Clasically Chaotic Systems}
\author[FaMAF]{F. M. Cucchietti},
\author[FaMAF]{H. M. Pastawski} and
\author[Stras]{R.Jalabert}
\address [FaMAF]{LANAIS de RMS, Facultad de Matem\'{a}tica, Astronom\'{i}a y F\'{i}sica, Universidad
Nacional de C\'{o}rdoba, C\'{o}rdoba, Argentina} 
\address [Stras]{Universit\'{e} Louis Pasteur, Strasbourg, France}

\begin{abstract}

The decay of overlap between a wave packet evolved with a Hamiltonian $\mathcal{H}$ 
and the same state  evolved with $\mathcal{H}+\Sigma $ serves as a measure of the 
decoherence time$\tau_{\phi}$. Recent experimental
and analytical evidence on classically chaotic systems suggest that, 
under certain conditions, 
 $\tau_{\phi}$ depends on  $\mathcal{H}$ but not on $\Sigma $.
 By solving numerically 
a Hamiltonian model we find evidence of
that property provided that the system shows a Wigner-Dyson spectrum 
(which defines quantum chaos) and the perturbation exceeds a crytical value 
defined by the parametric 
correlations of  the spectra.
\end{abstract}
\end{frontmatter}

The existence of chaos in classical mechanics is manifested in the evolution
of a state as an extreme sensitivity to the initial conditions. Quantum
mechanics, on the opposite, does not show this sensitivity.. This has raised
several problems in a dynamical definition of quantum chaos. In particular,
numerical \cite{Izrailev} and experimental\cite{Fink-letter} studies show
that time reversal can be achieved with great accuracy. Therefore, the
search for a quantum definition of chaos, lead to investigate the spectral
properties\cite{Bohigas} of quantum systems whose classical equivalent is
chaotic. Quantum chaos appears as the regime in which the properties of the
eigenstates follow the predictions of the Random Matrix Theory (RMT). In
particular the normalized spacing between energy levels $s=(\varepsilon
_{i+1}-\varepsilon _{i})/\Delta \varepsilon $, with $\Delta \varepsilon $
the mean level spacing, should have a probability distribution given by the
Wigner Dyson distribution $P_{WD}^{O}(s)=(\pi s/2)\ \exp (-\pi s^{2}/4)$ for
an orthogonal ensemble and $P_{WD}^{U}(s)=(32s^{2}/\pi ^{2})\ \exp
(-4s^{2}/\pi )$ for the unitary ensemble.

An infinite set of interacting spins is an example of a many-body system
which is chaotic in its classical version (lattice gas) and hence it is
expected to present the quantum signatures of chaos in the spectrum. The
dynamics of this particular system can be studied by Nuclear Magnetic
Resonance (NMR). Surprisingly, the ``diffusive'' dynamics of a local
excitation $\exp [-\mathrm{i}\mathcal{H}t_{R}]|0>$ can be reversed \cite
{polecho}, generating a Polarization Echo $M$ at time 2t$_{R}.$ In this case 
$\mathcal{H}$ is the many-body Hamiltonian of a network of spins with
dipolar interaction. To accomplish this, the transformation $\mathcal{H}%
\rightarrow -(\mathcal{H}+\Sigma )$ at time $t_{R}$ is performed with
standard NMR techniques \cite{polecho}. This transformation is possible due
to the anisotropic nature of the dipolar interaction. The perturbation $%
\Sigma $ is a non-invertible component of the Hamiltonian. In some systems,
the only contribution to $\Sigma $ is proportional to the inverse of the
radio frequency power and hence can be made arbitrarily small. In one body
systems the Polarization Echo (i.e. magnitude of the excitation recovered at
time $2t_{R})$ can then be written exactly as: 
\begin{equation}
M(t)=|\left\langle 0\right| \exp [\mathrm{i}(\mathcal{H}+\Sigma )t/\hbar
]\exp [-\mathrm{i}\mathcal{H}t/\hbar ]\left| 0\right\rangle |^{2},
\label{Overlap}
\end{equation}
where $\left| 0\right\rangle $ is the initial wave function, $\mathcal{H}$
the unperturbed Hamiltonian and $\Sigma $ the perturbation which can be
associated with an environmental disturbance. Then the magnitude of
experimental interest is the overlap between the same initial wave function
evolved with the two different Hamiltonians, $\mathcal{H}$ and $-(\mathcal{H}%
+\Sigma )$. We should note that the second evolution can be seen as an ``%
\textit{imperfect time reversal}'' of the wave function!. Consistently, more
than 10 years ago Peres \cite{Peres} proposed that dynamical signatures of
quantum chaos should be searched on the sensitivity to perturbations in the
Hamiltonian. Actually, for a classically chaotic system a perturbation in
the initial conditions is equivalent to a perturbation in the Hamiltonian.
The experiments show that $M$ decays rapidly with $t_{R}$ with Gaussian law 
\cite{levsteinpast} indicating a progressive failure in rebuilding the
original state. We can define a decoherence time $\tau _{\phi }$ from this
failure, as the width of the Gaussian. This is found\cite{Potencia} to be
roughly independent of $\Sigma $ and it extrapolates to a finite value when $%
\Sigma \rightarrow 0.$ Using a semiclassical one-body analytical approach in
classically chaotic systems characterized by a Lyapunov exponent $\lambda $,
Pastawski and Jalabert\cite{JalPast} have shown that there is a regime where
the attenuation of $M$ it is independent of the perturbation $\Sigma $ and
becoming $1/\tau _{\phi }=\lambda .$ This non-perturbative result is valid
for long times and as long as $\Sigma $ does not change the Hamiltonian
nature. Our general aim is to find numerical evidence of this regime where $%
M(t)$ is independent of $\Sigma $ considering the simplest Hamiltonian that
could model spin diffusion.

In this work, we study one-body Hamiltonian in quasi-1D systems with $N$
states which we called \textit{the Star's necklace model}. More
specifically, we use a tight binding model of a ring-shaped lattice with
on-site disorder, hopping matrix elements $V$, and a magnetic flux $\Phi $
perpendicular to the plane of the ring (see inset in Fig.1). Let us discuss
the general features through one representative class, each star has $20$
sites and there are $L=35$ beads in the necklace which makes $N=700$. In our
case, perturbation acts only between two star beads: $\Sigma (\delta \Phi
)=V\exp [\mathrm{i}2\pi \Phi /\Phi _{o}](\exp [\mathrm{i}2\pi \delta \Phi
/\Phi _{o}]-1)\left| 1\right\rangle \left\langle L\right| +c.c.$ Bra and ket
states contain orbitals in the star. The localized wave packet with energy $%
\left\langle 0\right| \mathcal{H}\left| 0\right\rangle \simeq 0$ moves along
the string and contains only $3N/4$ states. Anderson disorder is $W=3V$
which gives a $\tau _{\mathrm{imp}}\simeq 3\hbar /V$ and $\Phi =0.1\Phi _{0}.
$ We verify that for $\delta \Phi =0$ the dynamics of the system follows a
diffusive law, and that for all $\delta \Phi $ the statistics of the
eigenvalues correspond to those predicted by RMT (see Fig.1). We interpret
these fact as a signature of chaos. However, when studying the parametric
correlations of the energy spectrum\cite{Szafer} as a function of $\delta
\Phi $ the correlation function is definite positive. A critical value $%
\delta \Phi _{c}\sim 0.1$ $\Phi _{0}$ can be extrapolated from the strong
decay.

For small $t$ the decay of $M(t)$ is Gaussian like with a characteristic
time scaling with $\Sigma \,$exactly as could be predicted by perturbation
theory\cite{Peres}$.$ After a certain time of the order of the collision
time $\tau _{\mathrm{imp}}$ it becomes a stretched exponential with a
characteristic time $\tau _{\phi }$ independent of $\Sigma ,$ 
\begin{equation}
M(t)\sim \exp (-t/\tau _{\phi })^{\nu }+M_{\infty },  \label{stretch}
\end{equation} 
with $\nu \sim 0.87.$ and $\tau _{\phi }\sim 18\hbar /V$ (see Fig. 2). The
constant $M_{\infty }$ arises from finite size effects. Nonetheless, if the
perturbation does not exceed the critical threshold consistent with that of
the correlation function, $M_{\infty }$ increases with decreasing $\Sigma .$
On the other hand, if the perturbation is large $\delta \Phi > \delta
\Phi _{c},$ one gets $M_{\infty }\sim 1/N.$ (see inset of Fig. 2). The fact
that $\Sigma _{c}$ goes to zero when $N$ goes to infinity\cite{Szafer},
together with results of Fig. 2 for a finite system, could be a signature of
the existence of a nontrivial thermodynamic limit $\lim\limits_{\Sigma
\rightarrow 0}\lim\limits_{N\rightarrow \infty }M_{\infty }=0$ different
from the non-thermodynamic one $\lim\limits_{N\rightarrow \infty
}\lim\limits_{\Sigma \rightarrow 0}M_{\infty }=1$. Preliminary results on
the variation of $\tau _{\phi }$ with the amount of disorder and system size
are consistent with this hypothesis.

In order to present in a graphical way the physical phenomena of
decoherence, we calculated the weight of the wave functions evolved with the
two different Hamiltonians and the overlap between them as a function of the
layer. The results are shown in Fig. 3. It can be seen that the evolution of
the probabilities described by the perturbed and unperturbed Hamiltonians
are not significantly different, i.e. they would give the same coarse
grained values. However, the perturbation causes spatial fluctuations in the
phases which produce an integral overlap that decays to zero.

To sum up, our numerical calculations of the polarization echo $M$ in a
simple chaotic model indicate that there is a regime of the perturbation
where the decoherence time depends only on the perturbation. This basic
feature is also found in experiments\cite{Potencia} and in other theoretical
models. Some differences in the details remain, such as the value in the
exponent $\nu .$ According to preliminary numerical evidence $\nu $ could be
related to the particular topology \cite{diff-anom}induced by the matrix
elements in the Hamiltonian model. Different systems such as disordered
cylinders, maximally connected Hamiltonians (RMT) and chaotic stadiums
should be studied in order to characterize this dependence.

\begin{figure}[tbp]
\caption{Probability distribution of the normalized energy level spacing $s$
for the system studied. The black line is the Wigner-Dyson distribution for
the unitary ensemble, which is in reasonable agreement with the numerical
data. In the inset is shown an schematics of the Hamiltonian model, where
the system has the shape of a ring, and the in-layer sites are fully
connected.}
\end{figure}

\begin{figure}[tbp]
\caption{Plot of the decoherence time of the system according to \ref{stretch} 
as a function of $\delta \Phi $. The curve is for a system
with $20$ sites per layer and a diameter of $35$ sites, the amount of
disorder is $W=3$. The inset shows the dependence of the  asymptotic value 
$M_{\infty }$ with $\delta \Phi $. The straight line corresponds to $4/3N$.
As it can be seen $M_{\infty }$ reaches this value when $\delta \Phi > \delta%
\Phi _{c}$}
\end{figure}

\begin{figure}[tbp]
\caption{This figure shows that while the wave functions evolved with $%
\mathcal{H}$ and $\mathcal{H}+\Sigma $ have very similar probability
densities, the overlap between them decays rapidly to zero due to
interference effects. In thick solid line, profile of the initial wave
function and overlap of the two initial wave functions as a function of the
layer respectively. In thin solid lines, profile of the wave functions
evolved with two different Hamiltonians as a function of the layer. Thick
dotted line, overlap between the two evolved wave functions. The system has $%
15$ sites per layer and a diameter of $35$ sites, $W=3$, $t=2 \hbar/V$, and
$\delta \Phi=0.1$ $\Phi _{0}.$ The sum over layers for this overlap is equal
to $1.4$ $10^{-2}$}
\end{figure}

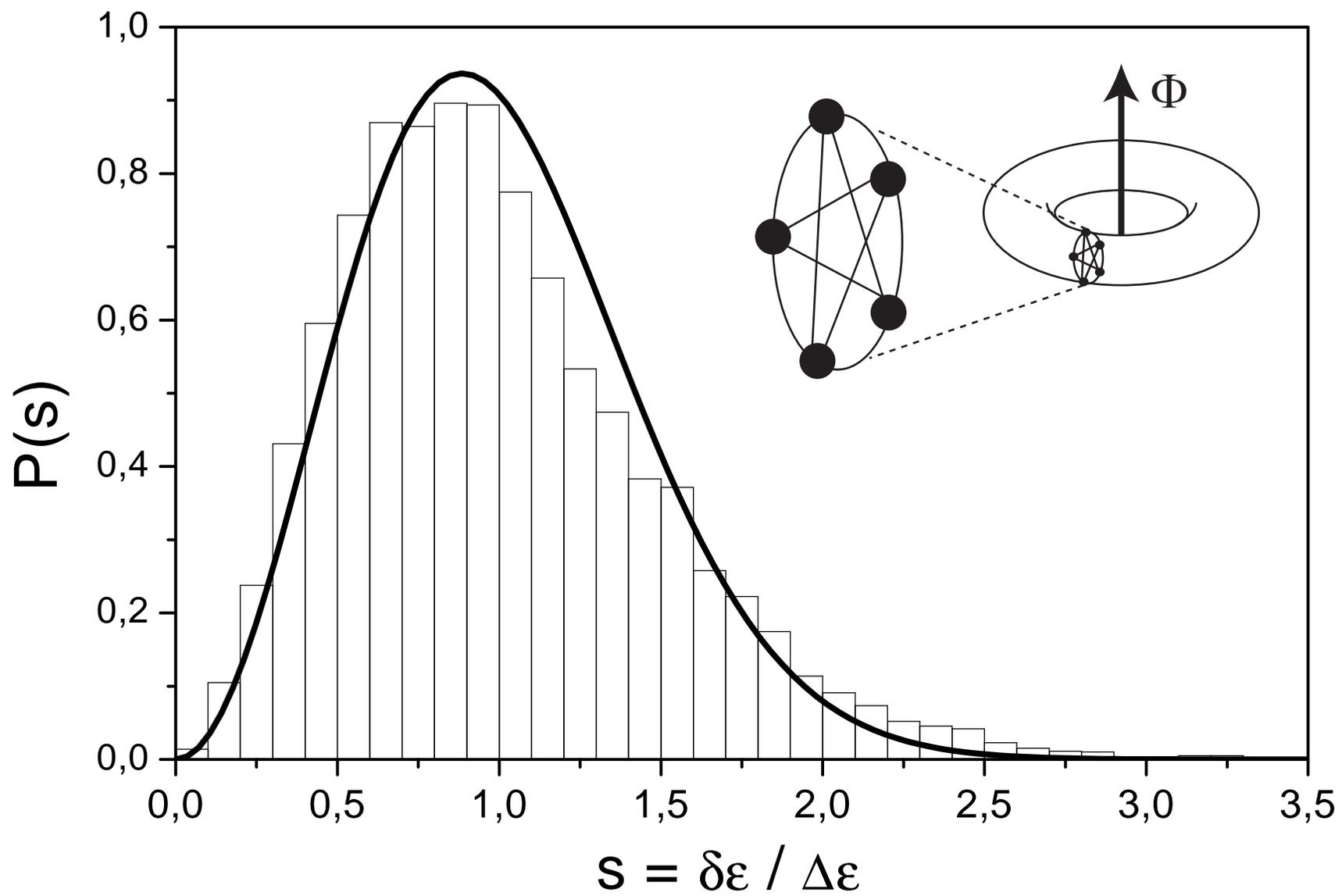
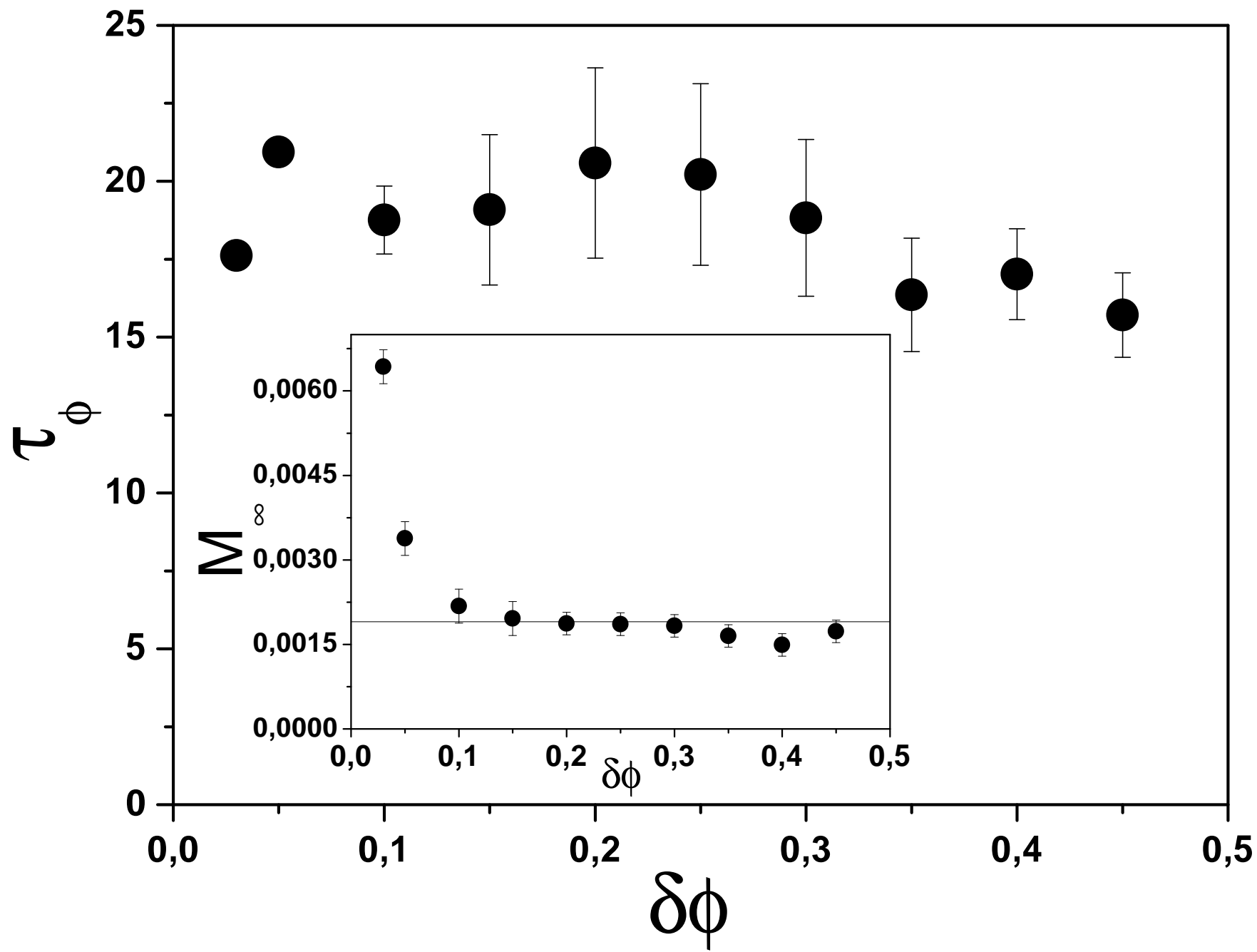
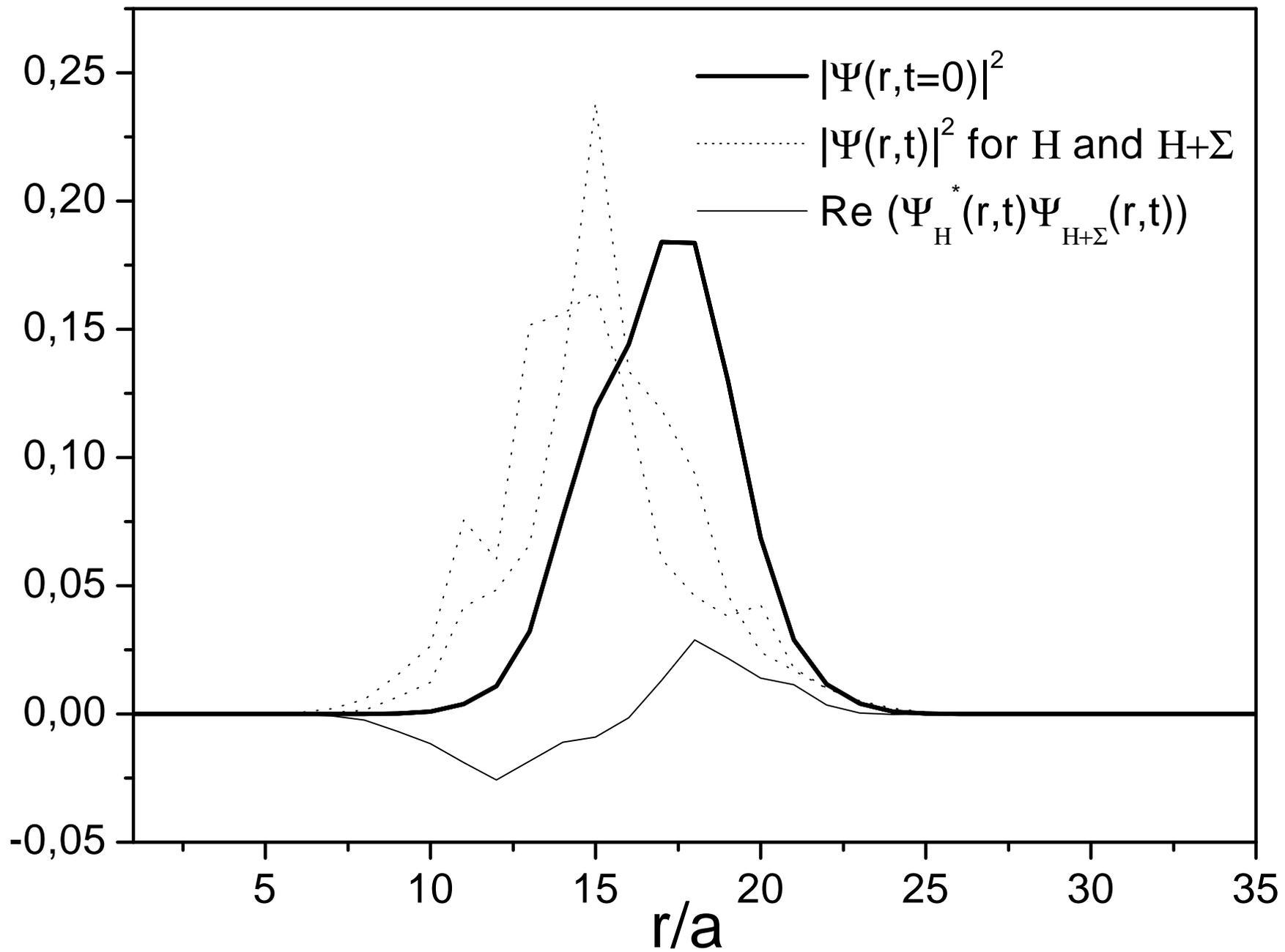
\end{document}